\documentclass[11pt,eqs]{article}

\usepackage{latexsym,amsmath,amstext,amssymb,mathtools}
\usepackage{graphicx}
\usepackage[all]{xy}

\def\cleq{\setcounter{equation}{0}}

\usepackage[margin=1in]{geometry}

\title{
Open string T-duality in a weakly curved background
}
\author{Lj. Davidovi\'c\thanks{e-mail: ljubica@ipb.ac.rs}\\
{\it Institute of Physics,}
{\it University of Belgrade,}
{\it 11080 Belgrade, Serbia}
}

\begin{document}
\maketitle

\begin{abstract}
We consider a theory of an open string
moving in a weakly curved background,
composed of a constant metric and a linearly coordinate dependent Kalb-Ramond field with an infinitesimal field strength.
We find its T-dual using the generalized Buscher procedure developed for the closed string moving in a
weakly curved background,
and the fact that
solving the bo\-u\-nda\-ry conditions, the open string theory transforms to
the effective closed string theory.
So, T-dualizing the
effective theory along all effective directions we
obtain its T-dual theory and 
resume the open string theory which has such an effective theory.
In this way we obtain the open string theory T-dual.
\end{abstract}

\section{Introduction}

T-duality (reviewed in \cite{GPR,AGL}) is a symmetry of a string spectrum, 
exchanging the momentum and the winding numbers,
a symmetry which
was not encountered in any point particle theory \cite{GPR,AGL,ZW,BBS},
a symmetry which is 
therefore naturally assumed to be connected with the fact that 
the strings unlike the point particles can wrap around compactified dimensions \cite{BGS,KY,SS}.
T-duality is a symmetry which connects string theories
which are physically equivalent but describe strings which move in space-times with different geometries of the compactified dimensions.
The theory with one dimension 
compactified on a circle of
radius R is by T-duality physically equivalent to the theory with one dimension compactified  on a circle of radius $l_{s}^{2}/R$, where $l_{s}$  is the
fundamental string length scale.

The existence of T-duality initiated the search for the T-dualization procedure,
the set of rules which give a prescription for determining all string theories T-dual to the one given theory.
The problem was solved for some particular backgrounds. 
The first investigations addressing the problem
were done for a
string sigma model describing a string moving in a
 background containing a metric $G_{\mu\nu}$, an antisymmetric field $B_{\mu\nu}$ and a dilaton field $\Phi$,
and it was required that the metric admits at least one abelian isometry which leaves the action invariant. The investigation resulted
in the Buscher procedure \cite{B,BB,RV}. 
This procedure offers the prescription, founded in localizing the global isometry.
The main object of the procedure is a gauge invariant action,
which carries the information on both the initial theory and the T-dual theory,
so that both of these theories can be obtained from the qauge invariant action
for one of the two of its equations of motion.
The procedure is applicable to the constant background and the coordinate dependent backgrounds, but only 
along directions which do not appear as the background field's arguments.

It was found that in a T-dualization of a background along isometry direction
one mixes the background fields $G_{\mu\nu}$ and $B_{\mu\nu}$,
which explains how T-duality relates backgrounds with different geometrical and topological properties.
The search for the equivalent string backgrounds goes beyond the Buscher procedure.
If one considers a string sigma model as a theory of maps from two dimensional space to a manifold $M$, the geometrical string background are obtained if the transition functions in the overlapping coordinate patches are diffeomorphisms and gauge transformations. If T-duality is included into transition functions one obtains non-geometrical string backgrounds \cite{HULL}. Even more non-geometrical backgrounds are known to exist \cite{DH}, which in some cases arise from the consistent string backgrounds performing generalized T-dualities in non-isometry directions.

T-dualization procedures were developed for the clo\-sed strings
and then extended to the open string case.
T-duality of an open string differs from the closed string T-duality,
due to a simple fact that the open string does not have a winding number.
Its ends move on the Dirichlet branes and can simply unwind \cite{P}.
It was shown in \cite{Pol} that Dirichlet branes have a significant role in string duality.
The open string T-duality along the isometry directions was investigated using
the standard Buscher procedure \cite{BL}, canonical transformations \cite{BL,DO},
the functional integral approach \cite{DO} and using new formalisms for nonabelian isometry \cite{DO,FKS,OQ}.
T-duality relates open strings to  Dirichlet branes, which is a limiting case of duality
between pairs of  Dirichlet branes \cite{KS}.
Poisson-Lie T-duality enabled the investigation of backgrounds without isometries.

References \cite{DS3,DS4,DNS} offered the generalized Buscher T-du\-a\-li\-za\-tion procedure,
which is applicable along an arbitrary coordinate regardless of whether this coordinate appears as an argument of the background fields or not.
The main difference between the standard and the generalized procedure is that,
in the generalized procedure, in addition to the introduction of
the gauge fields and the covariant derivatives one should also introduce an invariant argument which is defined as a line integral of
the covariant derivatives of the original argument.
The generalized procedure was realized for
the closed string moving in a curved background with an infinitesimal coordinate dependence, named the weakly curved background  \cite{DS3}.
A more general procedure was defined in  \cite{{DS4}}
for a weakly curved background of 
the second order, which does not possess the global isometry.
In this paper,
we will use the first version of the generalized  T-dualization procedure to address the T-duality of an open string moving in the weakly curved background.

The investigation of an open bosonic string moving in a
weakly curved background,
presented  in Refs. \cite{DS,DS1,DS2},
dedicated to solving the boundary conditions,
showed that the effective theory obtained for the solution of the boundary conditions
is in fact a closed string theory.
The boundary conditions were treated by Dirac method in \cite{DS,DS1}.
Reference \cite{DS2} included a close examination of the parity of the
boundary conditions, equations of motion and the consistency conditions.
The two approaches lead to equivalent solutions.
The effective theory, obtained for the solution, is defined on the doubled target
space, which consists of a symmetric and an antisymmetric variables $(q^\mu, \tilde{q}^\mu)$,
with the first variable being the even part of the initial coordinate and the second variable being the double of the first variable.
The effective metric depends on the effective coordinate $q^\mu$, while
the effective Kalb-Ramond field depends on a double variable ${\tilde{q}}^\nu$.

In this paper, we will apply the generalized Buscher procedure \cite{DS3,DS4,DNS},
developed for the closed string moving in the weakly curved background,
to the effective closed string theory, obtained for the solution of the open string boundary conditions,
along all effective coordinates.
We will obtain its T-dual theory.
We will show that T-dualization of the obtained theory along all dual coordinates leads to the initial theory. 
Our main goal is to find the open string T-dual.
So, we'll assume that the obtained T-dual theory is an effective theory of some open string theory. We search for the explicit form of this theory.
Demanding that the effective theory of the unknown theory is exactly the T-dual theory, we obtain a T-dual of the initial open string theory. 
The obtained theory is defined in the geometrical space, on the contrary to the closed string case, where the T-dualization led to a T-dual theory
defined on the non-geometrical double space.
The relations between the initial background and its T-dual differ from those in the closed string case.


\section{Open string theory in a weakly curved background}

Let us consider an open bosonic string moving
in a background defined by a metric tensor
$G_{\mu\nu}$, an antisymmetric Kalb-Ramond
field $B_{\mu\nu}$,
and a dilaton field $\Phi$, described by the 
action  \cite{ZW,BBS}
\begin{eqnarray}\label{eq:action0}
S[x]= \kappa \int_{\Sigma} d^2\xi\sqrt{-g} \Big[\Big(\frac{1}{2}{g}^{\alpha\beta}G_{\mu\nu}(x)
+\frac{\varepsilon^{\alpha\beta}}{\sqrt{-g}}B_{\mu\nu}(x)\Big)
\partial_{\alpha}x^{\mu}\partial_{\beta}x^{\nu}+\frac{1}{4\pi\kappa}\Phi(x)R^{(2)}\Big],
\end{eqnarray}
where
the integration goes over a 2-dimensional world-sheet $\Sigma$
parametrized by
$\xi^\alpha$ ($\xi^{0}=\tau,\ \xi^{1}=\sigma$),
$g_{\alpha\beta}$ is the intrinsic world-sheet metric, $R^{(2)}$ corresponding 2-di\-me\-nsi\-o\-nal scalar curvature,
$x^{\mu}(\xi),\ \mu=0,1,...,D-1$ are the coordinates of the
D-dimensional space-time,
$\kappa=\frac{1}{2\pi\alpha^\prime}$
with $\alpha^\prime$ being the Regge slope parameter
and $\varepsilon^{01}=-1$. We will also use the notation
$\dot{x}=\frac{\partial x}{\partial\tau}$,
$x^\prime=\frac{\partial x}{\partial\sigma}$.
In order to have a conformal invariance on a quantum level the background fields $G_{\mu\nu}, B_{\mu\nu},$ and $\Phi$ have to obey 
the following space-time equations of motion
( given in the lowest order in the slope parameter $\alpha^\prime$)
\begin{eqnarray}\label{eq:beta}
&&
R_{\mu\nu}-\frac{1}{4}B_{\mu\rho\sigma}B_\nu^{\ \rho\sigma}+2D_\mu \partial_\nu\Phi=0,
\nonumber\\
&&
D_\rho B^\rho_{\ \mu\nu}-2\partial_\rho\Phi B^\rho_{\ \mu\nu}=0,
\nonumber\\
&&
4(\partial\Phi)^{2}-4D_\mu\partial^\mu\Phi
+\frac{B_{\mu\nu\rho}B^{\mu\nu\rho}}{12}-R
+4\pi\kappa\frac{D-26}{3}
=0\,,\nonumber\\
\end{eqnarray}
where
$B_{\mu\nu\rho}=\partial_\mu B_{\nu\rho}
+\partial_\nu B_{\rho\mu}+\partial_\rho B_{\mu\nu}$
is the field strength of the field $B_{\mu \nu}$, and
$R_{\mu \nu}$ and $D_\mu$ are the Ricci tensor and the
covariant derivative with respect to the space-time metric.
We will consider the open string moving in the weakly curved background \cite{VS,CSS,HKK,ARS},
defined by
\begin{eqnarray}\label{eq:gb}
G_{\mu\nu}&=&const,
\nonumber\\
B_{\mu\nu}(x)&=&b_{\mu\nu}+h_{\mu\nu}(x),\quad h_{\mu\nu}(x)=\frac{1}{3}B_{\mu\nu\rho}x^\rho,
\nonumber\\
\Phi&=&const,
\end{eqnarray}
where the quantities $b_{\mu\nu}$ and $B_{\mu\nu\rho}$ are constant.
For such a background the equations of motion (\ref{eq:beta}) reduce to
\begin{eqnarray}\label{eq:betared}
&&R_{\mu \nu} - \frac{1}{4} B_{\mu \rho \sigma}
B_{\nu}^{\ \rho \sigma}=0\, ,
\nonumber\\
&&D_\rho B^{\rho}_{\ \mu \nu} = 0,
\nonumber\\
&&\frac{1}{12}{B_{\mu\nu\rho}B^{\mu\nu\rho}}-R
+4\pi\kappa\frac{D-26}{3}
=0.
\end{eqnarray}
The weakly curved background is the solution of these equations for $D=26$, if  the quadratic terms in
$B_{\mu\nu\rho}$ can be neglected. So, we choose $B_{\mu\nu\rho}$  infinitesimal and
work in the linear order in  $B_{\mu\nu\rho}$.

The action (\ref{eq:action0}) for the string moving in the weakly curved background (\ref{eq:gb}) can be written as 
\begin{equation}\label{eq:action1}
S[x] = \kappa \int_{\Sigma} d^2\xi\
\partial_{+}x^{\mu}
\Pi_{+\mu\nu}(x)
\partial_{-}x^{\nu},
\end{equation}
where the light-cone coordinates and their derivatives are
$
\xi^{\pm}=\frac{1}{2}(\tau\pm\sigma),
\quad
\partial_{\pm}=
\partial_{\tau}\pm\partial_{\sigma},
$
and 
$\Pi_{\pm\mu\nu}$ is 
a combination of the background fields, defined by
\begin{eqnarray}\label{eq:pi}
\Pi_{\pm\mu\nu}(x)=
B_{\mu\nu}(x)
\pm\frac{1}{2}G_{\mu\nu}.
\end{eqnarray}
There is a direction of research in differential geometry
in which the above theory is interpreted as a theory of maps $x^\mu$ from the two-dimensional space $\Sigma$
to a manifold $M$ with metric $G$ and a closed $3$-form $H$ \cite{Hll}.
For a geometric string background,
the background fields in overlaps of the coordinate patches are related by diffeomorphisms and gauge transformations. The local choices fit together to form a space-time manifold.

The minimal action principle for the
open string, described by (\ref{eq:action1}), gives the equation of motion
\begin{equation}\label{eq:motion1}
\partial_{+}\partial_{-}x^\mu
-B^{\mu}_{\ \nu\rho}
\partial_{+}x^\nu\partial_{-}x^\rho=0,
\end{equation}
and the boundary conditions
on the string endpoints
\begin{equation}\label{eq:bonc}
\gamma^{0}_\mu\Big{|}_{\sigma=0,\pi}=0,
\end{equation}
where
\begin{eqnarray}\label{eq:bc}
\gamma^{0}_{\mu}\equiv \frac{\delta {\cal{L}}}{\delta
x^{\prime\mu}} 
=G_{\mu\nu}x^{\prime\nu}-2B_{\mu\nu}\dot{x}^\nu
=\Pi_{+\mu\nu}\partial_{-}x^\nu
+\Pi_{-\mu\nu}\partial_{+}x^\nu.
\end{eqnarray}

Solving of these boundary conditions was a subject of investigation of Refs. \cite{DS,DS1,DS2}.
In the first two papers, the boundary conditions were treated as constraints
and we applied the Dirac consistency procedure.
We obtained the infinite number of constraints, gathered them into
two parameter dependent con\-stra\-ints, which were solved.
We obtained the form of the initial coordinates satisfying the boundary conditions.
In the paper \cite{DS2}, we obtained the analogous result by examining the
parity of the equations of motion, boundary conditions and the consistency conditions.
The solution of the boundary condition at $\sigma=0$ is
\begin{eqnarray}\label{eq:stcoor}
\partial_\pm x^{\mu}&=&(G^{-1})^{\mu\rho}\Big[G_{\rho\nu} -A_{\rho\nu} ({\tilde{q}})
\pm2B_{\rho\nu}(q)
\Big]
\partial_\pm {q}^\nu,
\end{eqnarray}
where
$$
A_{\rho\nu}({{\tilde{q}}})
=
\Big{[}
h({{\tilde{q}}})-12bh({{\tilde{q}}})b
-12h(b{{\tilde{q}}})b+12bh(b{{\tilde{q}}})
\Big{]}_{\rho\nu},$$
$q^\mu$  is
the even part of the initial coordinate
\begin{eqnarray}\label{eq:qqbar}
q^\mu(\sigma)&=&
\sum_{n=0}^{\infty}\frac{{\sigma}^{2n}}{(2n)!}x^{(2n)\mu}\Big{|}_{\sigma=0},
\end{eqnarray}
and $\tilde{q}$ is  its double, which satisfies
\begin{eqnarray}\label{eq:dotprime}
{\dot{\tilde{q}}}^\mu= q^{\prime\mu} , \qquad
{\tilde{q}}^{\prime\mu}={\dot{q}}^\mu.
\end{eqnarray}

Changing the domain to $\sigma \in [-\pi,\pi]$,
one enables the solution (\ref{eq:stcoor}) to satisfy the boundary condition at the other string endpoint as well. For details see \cite{DS2}.
Substituting the solution (\ref{eq:stcoor}) to the initial action 
we obtain the effective theory, given in terms of effective variables $(q^\mu,\tilde{q}^\mu)$
\begin{equation}\label{eq:aceff}
S^{eff} =\kappa \int d\tau \int_{-\pi}^\pi d\sigma\, 
\partial_{+}q^{\mu}\,
\Pi^{eff}_{+\mu\nu}(q,2b\tilde{q})\,
\partial_{-}q^{\nu},
\end{equation}
with an effective background 
\begin{equation}\label{eq:effpi}
\Pi^{eff}_{\pm\mu\nu}(q,2b\tilde{q})\equiv
B^{eff}_{\mu\nu}(2b{\tilde{q}})
\pm
\frac{1}{2}G^{eff}_{\mu\nu}(q).
\end{equation}
The effective theory is defined on the doubled space $(q^\mu,\tilde{q}^\mu)$,
with a double coordinate appearing in a solution of the boundary conditions.
Usually,
the doubled geometry is introduced in an investigation of T-duality.
A coordinate and its double give an origin to
the winding and the momentum number, the exchange of which by T-duality does not change the physics.
The doubled description \cite{DF,AT} makes T-duality a manifest symmetry.

The effective metric and the Kalb-Ramond field in (\ref{eq:effpi}) are explicitly given by
\begin{eqnarray}\label{eq:effpolja}
G^{eff}_{\mu\nu}(q)&=& G^{E}_{\mu\nu}(q)\,
:=\big(G-4B^{2}(q)\big)_{\mu\nu},
\nonumber\\
B^{eff}_{\mu\nu}(2b{\tilde{q}})&=& -\frac{\kappa}{2}\Big(g_{E} \Delta\theta (2b{\tilde{q}})g_{E}\Big)_{\mu\nu},
\end{eqnarray}
where 
$\Delta\theta^{\mu\nu}$ is the infinitesimal part of
the non-co\-mmu\-ta\-ti\-vi\-ty parameter
\begin{eqnarray}\label{eq:effnon}
\theta^{\mu\nu}=-\frac{2}{\kappa}\Big{[}G^{-1}_{E}BG^{-1}\Big{]}^{\mu\nu}
=\theta_{0}^{\mu\nu}
-\frac{2}{\kappa}\Big{[}g_{E}^{-1}(h+4bhb)g_{E}^{-1}\Big{]}^{\mu\nu},
\end{eqnarray}
defined in analogy with the flat space-time non-co\-mmu\-ta\-ti\-vi\-ty parameter introduced
in \cite{SW}.
One can show that $\theta^{\mu\nu}$ is indeed the non-commutativity parameter by considering
the phase spaces 
of the initial and the effective theory,
as in \cite{DS}.
The initial phase space consists of the initial coordinates $x^\mu$ and the momenta $\pi_\mu$.
Solving the boundary conditions one obtains expressions for $x^\mu$ and $\pi_\mu$ given in terms of the effective coordinates $q^\mu$ and the effective momenta $p_\mu$ (the even part of the initial momenta). So, the initial phase space
transforms to the effective phase space, whose variables
satisfy the star brackets
\begin{equation}\label{eq:stbrac}
{}^{\star}\{q^\mu(\sigma),p_\nu(\bar{\sigma})\}
=2\delta^{\mu}_{\nu}\delta_{S}(\sigma,\bar{\sigma}),
\end{equation}
defined in Appendix B of Ref. \cite{DS},
which are the analogs of the Dirac brackets of the initial space.
As the initial coordinates which solve the boundary conditions depend on both 
effective coordinates and the effective momenta, they are non-commutative.
Taking 
\begin{equation}\label{eq:xcm}
X^\mu(\sigma)=x^\mu(\sigma)-x^{\mu}_{cm},\quad x^{\mu}_{cm}=\frac{1}{\pi}\int_{0}^{\pi}d\sigma x^\mu(\sigma),
\end{equation}
one finds that the non-commutativity parameter is exactly
$\theta^{\mu\nu}$ given by (\ref{eq:effnon})
\begin{eqnarray}\label{eq:xxx}
{}^{\star}\{X^\mu(\sigma),X^\nu(\bar{\sigma})\}
&=&\theta^{\mu\nu}[X(\sigma)]
\left\{ \begin{array}{rcl}
-1 && \sigma,\bar{\sigma}=0\\
1 && \sigma,\bar{\sigma}=\pi\\
0 && otherwise
\end{array}\right..
\end{eqnarray}

\section{T-dualization of the effective theory}
\cleq

In this section we will 
apply the generalized Buscher  T-dualization pocedure, developed in
\cite{DS3,DS4,DNS}, to the effective theory (\ref{eq:aceff})
and find its T-dual.
In general, the procedure could be applied along arbitrary set of the effective coordinates.
In this paper, we will investigate only the application along all effective coordinates.
The effective theory is given in terms of the effective coordinate $q^\mu$ and its double $\tilde{q}^\mu$,
so the overall procedure will look like the T-dualization of the T-dual of the closed string theory in the weakly curved background,
because the weakly curved background T-dual is also defined in doubled space, composed of a dual coordinate (Lagrange multiplier) and its double.

So, let us apply the generalized Buscher T-du\-ali\-za\-ti\-on procedure to the effective theory (\ref{eq:aceff}) along all effective directions $q^\mu$.
The first task is to localize the global symmetry $\delta q^\mu=\lambda^\mu=const$ and find the gauge invariant action.
Following the procedure, we substitute 
the light-cone derivatives $\partial_\pm q^\mu$ of the effective theory (\ref{eq:aceff}) with the covariant derivatives $D_\pm q^\mu$, defined by
\begin{equation}
D_\pm q^\mu=\partial_\pm q^\mu+v^\mu_\pm,
\end{equation}
where $v^\mu_\pm$ are the gauge fields, which transform as $\delta v^\mu_\pm=-\partial_\pm\lambda^\mu$.
We also substitute the argument of the background fields, with an invariant argument,
which is obtained substituting the effective coordinate $q^\mu$ and its double $\tilde{q}^\mu$ with
an invariant effective coordinate and its double, defined
as the line integrals of the covariant derivatives of the effective coordinate and its double,
with the line integrals taken along the path P, from the initial point $\xi^\alpha_{0}$
to the point $\xi^\alpha$.
The last step in forming the gauge invariant action is adding the
Lagrange multiplier term $\frac{1}{2}(v^\mu_{+}\partial_{-}\varrho_\mu-v^\mu_{-}\partial_{+}\varrho_\mu)$,
which makes the introduced gauge fields nonphysical.
Fixing the gauge by $q^\mu(\xi)=q^\mu(\xi_{0})$,
we obtain the gauge fixed action, which reads
\begin{eqnarray}\label{eq:pomocno}
{\cal S}_{fix}=\kappa\int d^{2}\xi\Big[
v^\mu_{+}
\Pi^{eff}_{+\mu\nu}(\Delta V,2b\Delta \tilde{V})
v^\nu_{-}
+
\frac{1}{2}(v^\mu_{+}\partial_{-}\varrho_\mu-v^\mu_{-}\partial_{+}\varrho_\mu)
\Big],
\end{eqnarray}
where $\Delta V^\mu$ and $\Delta\tilde{V}^\mu$
are the following line integrals of the gauge fields:
\begin{eqnarray}\label{eq:vvtilda}
\Delta V^\mu&=&\int_{P}(d\xi^{+} v^{\mu}_{+}
+d\xi^{-} v^{\mu}_{-}),
\nonumber\\
\Delta\tilde{V}^\mu&=&\int_{P}(d\xi^{+} v^{\mu}_{+}
-d\xi^{-} v^{\mu}_{-}).
\end{eqnarray}

The next step in the T-dualization procedure is finding the equations of motion of the gauge fixed action.
For the equation of motion for the Lagrange multiplier the action will reduce to the initial action, while for the equation for the gauge fields the action will become the T-dual action. 
Varying the action (\ref{eq:pomocno}), one obtains the following equations of motion:
\begin{equation}\label{eq:jed}
\partial_{-}v^\mu_{+}-\partial_{+}v^\mu_{-}=0,
\end{equation}
and
\begin{equation}\label{eq:eqm}
\Pi^{eff}_{\pm\mu\nu}
(\Delta V,2b\Delta \tilde{V})
v^\nu_\mp
+\frac{1}{2}\partial_\mp\varrho_\mu=\pm\beta^\pm_\mu(V),
\end{equation}
where the beta functions $\beta^\pm_\mu(V)$ come from the variation over the background field's arguments, and they are defined by
\begin{eqnarray}
 \delta_{A}{\cal S}_{fix}&=&
\kappa\int d\xi^{2}\Big[\varepsilon^{\alpha\beta}\partial_\rho B^{eff}_{\mu\nu}(2b\delta\tilde{V})^\rho
+\frac{1}{2}\eta^{\alpha\beta}\partial_\rho G^{eff}_{\mu\nu}\delta V^\rho\Big]
\partial_\alpha V^\mu\partial_\beta V^\nu
\nonumber\\
&=&-\kappa\int d^{2}\xi(\beta^{+}_\mu\delta v^\mu_{+}+\beta^{-}_\mu\delta v^\mu_{-}).
\end{eqnarray}
Partially integrating, using the fact that $\partial_\alpha\delta \tilde{V}^\mu=\varepsilon^\beta_{\ \alpha}\delta v^\mu_\beta$, one obtains the explicit form of
the beta functions
\begin{equation}\label{eq:bete}
\beta^\pm_\mu(V)=\Big[
-b_\mu^{\ \sigma}\partial_\sigma B^{eff}_{\nu\rho}
+\frac{1}{4}\partial_\mu G^{eff}_{\nu\rho}
\Big]V^\nu\partial_\mp V^\rho.
\end{equation}
Substituting the explicit values of the effective background fields (\ref{eq:effpolja}), (\ref{eq:emet}) and (\ref{eq:effnon}) one obtains
\begin{eqnarray}
\beta^\pm_\mu(V)&=&-\frac{1}{3}\Big[ 
b_\mu^{\ \sigma}B_{\sigma\nu\rho}+
b_\nu^{\ \sigma}B_{\sigma\rho\mu}
+b_\rho^{\ \sigma}B_{\sigma\nu\mu}
-4b_\mu^{\ \sigma} b_\nu^{\ \tau} b_\rho^{\ \varepsilon}B_{\sigma\tau\varepsilon}
\Big]V^\nu\partial_\mp V^\rho.
\end{eqnarray}
\subsection{Regaining the effective theory}

Solving Eq. 
(\ref{eq:jed}), by 
\begin{equation}\label{eq:pz}
v^\mu_\pm=\partial_\pm q^\mu,
\end{equation}
one obtains
$V^\mu=q^\mu$ and $\tilde{V}^\mu=\tilde{q}^\mu$.
Substituting these relations to the gauge fixed action
(\ref{eq:pomocno}), one confirms that it reduces to the effective action (\ref{eq:aceff}).

\subsection{T-dual theory}

To obtain the T-dual action,
one should substitute the expressions for the gauge fields to the gauge fixed action.
The gauge fields are obtained, multiplying 
 the equation of motion (\ref{eq:eqm}),
by the inverse of $\Pi^{eff}_{\pm\mu\nu}$,
$(\Theta^{eff}_\mp)^{\mu\nu}$ defined in (\ref{eq:tetaeff})
\begin{equation}\label{eq:dz}
v_\pm^\mu=-\kappa(\Theta^{eff}_\pm)^{\mu\nu}
(\Delta V,2b\Delta \tilde{V})
\Big[
\partial_\pm\varrho_\nu\pm2\beta^\mp_\nu(V)
\Big].
\end{equation}
The beta functions will not contribute to the T-dual action because they are infinitesimal and appear within the quadratic term. So, the
T-dual action reads
\begin{equation}\label{eq:tdualact}
^\star\!{\cal S}=\kappa\int\, d^{2}\xi
\partial_{+}\varrho_\mu
\frac{\kappa}{2}(\Theta^{eff}_{-})^{\mu\nu}
(\Delta V(\varrho),2b\Delta \tilde{V}(\varrho))
\partial_{-}\varrho_\nu.
\end{equation}

Let us calculate the argument of the background fields.
Using the zeroth order of the  equations of motion (\ref{eq:eqm}) and (\ref{eq:pieff})
\begin{equation}
\Pi^{eff}_{0\pm\mu\nu}
v^\nu_{0\mp}
+\frac{1}{2}\partial_\mp\varrho_\mu=
\pm\frac{1}{2}g^{eff}_{\mu\nu}
v^\nu_{0\mp}
+\frac{1}{2}\partial_\mp\varrho_\mu=0,
\end{equation}
and the fact that the open string effective metric (\ref{eq:emet}) is the same in both initial and the effective background (\ref{eq:effmet}),
one obtains the explicit value of the arguments (\ref{eq:vvtilda}),
\begin{eqnarray}
&&V^\mu_{0}(\varrho)=
(g_{E}^{-1})^{\mu\nu}(G^{eff},B^{eff})\tilde\varrho_\nu=
(g_{E}^{-1})^{\mu\nu}\tilde\varrho_\nu
,
\nonumber\\
&&\tilde{V}^\mu_{0}(\varrho)=(g_{E}^{-1})^{\mu\nu}(G^{eff},B^{eff})\varrho_\nu
=(g_{E}^{-1})^{\mu\nu}\varrho_\nu.
\end{eqnarray}

Comparing the forms of the effective action (\ref{eq:aceff}) and the T-dual action (\ref{eq:tdualact}), we see that they are equal under the following transformations
\begin{eqnarray}\label{eq:up}
\partial_\pm q^\mu&\rightarrow&\partial_\pm\varrho_\mu,
\nonumber\\
\Pi^{eff}_{+\mu\nu}&\rightarrow&^\star\Pi_{+}^{\mu\nu},
\end{eqnarray}
where the T-dual background is
\begin{equation}
^\star\Pi_{+}^{\mu\nu}(\Delta V,2b\Delta \tilde{V}):=\frac{\kappa}{2}(\Theta^{eff}_{-})^{\mu\nu}(\Delta V,2b\Delta \tilde{V}).
\end{equation}
Using (\ref{eq:tetaeff}), we obtain the T-dual metric $^\star{G}^{\mu\nu}$, which depends on the first variable $\Delta V^\mu$ and the T-dual Kalb-Ramond field $^\star{B}^{\mu\nu}$,
 which depends on the second variable $2b^\mu_{\ \nu}\Delta \tilde{V}^\nu$
\begin{eqnarray}\label{eq:dualnapolja}
^\star{G}^{\mu\nu}&=&(G_{E}^{-1})^{\mu\nu}(\Delta V),
\nonumber\\
^\star{B}^{\mu\nu}&=&\frac{\kappa}{2}(\theta^{eff})^{\mu\nu}(2b\Delta \tilde{V})=
\frac{\kappa}{2}\Delta\theta^{\mu\nu}(2b\Delta \tilde{V}).
\end{eqnarray}
We see that the effective metric has transformed to its inverse and that the Kalb-Ramond field has transformed to the infinitesimal part of the non-co\-mmu\-ta\-ti\-vi\-ty parameter.

Comparing the actions one could conclude that the relation between
the variables of the effective theory and its T-dual 
is simple as in (\ref{eq:up}).
However, the real connection is given by the T-dual coordinate transformation law,
which is obtained comparing the expressions for the
gauge fields (\ref{eq:pz}) and (\ref{eq:dz}), and it reads
\begin{eqnarray}\label{eq:zakon}
&&\partial_\pm q^\mu\cong-\kappa(\Theta^{eff}_\pm)^{\mu\nu}(\Delta V(\varrho),2b\Delta \tilde{V}(\varrho))
\Big[
\partial_\pm\varrho_\nu\pm2\beta^\mp_\nu(V(\varrho))
\Big].
\end{eqnarray}
In the zeroth order this law gives
\begin{equation}
q^{(0)\mu}\cong V^{(0)\mu}(\varrho),
\end{equation}
which will be useful later on.
\section{T-dulization of T-dual theory}
\cleq
Let us now show that the T-dualization of the T-dual theory (\ref{eq:tdualact}) leads to the initial effective theory (\ref{eq:aceff}).
Following the T-dualization procedure, we localize the global symmetry $\delta\varrho_\mu=\lambda_\mu=const$.
We introduce the gauge fields $u_{\pm\mu}$, which transform as
$\delta u_{\pm\mu}=-\partial_\pm\lambda_\mu$,
 substitute the ordinary derivatives $\partial_\pm\varrho_\mu$ in  the T-dual action (\ref{eq:tdualact}) by the covariant derivatives
$D_\pm\varrho_\mu=\partial_\pm\varrho_\mu+u_{\pm\mu}$, substitute the coordinate $\varrho_\mu$ and its double $\tilde\varrho_\mu$
in the background field's argument by an invariant coordinate 
$\varrho^{inv}_\mu=\varrho_\mu(\xi)-\varrho_\mu(\xi_{0})+\Delta U_\mu$
and its double $\tilde\varrho^{inv}_\mu=\tilde\varrho_\mu(\xi)-\tilde\varrho_\mu(\xi_{0})+\Delta\widetilde{U}_\mu$
where
$\Delta U_\mu=\int_{P}(d\xi^{+} u_{+\mu}
+d\xi^{-} u_{-\mu}),
$ and
$\Delta\widetilde{U}_\mu=\int_{P}(d\xi^{+} u_{+\mu}
-d\xi^{-} u_{-\mu})$,
add the Lagrange multiplier $\varsigma^\mu$ term and fix the gauge by $\varrho_\mu(\xi)=\varrho_\mu(\xi_{0})$. In this way, we obtain
the gauge fixed action
for the T-dual action (\ref{eq:tdualact}), which reads
\begin{eqnarray}\label{eq:tdualactpom}
^\star{\cal S}_{fix}&=&\kappa\int d^{2}\xi\Big[
\frac{\kappa}{2}(\Theta^{eff}_{-})^{\mu\nu}
\big(\Delta V(\Delta U),2b\Delta \tilde{V}(\Delta U)\big)
u_{+\mu}u_{-\nu}
+\frac{1}{2}\big(u_{+\mu}\partial_{-}\varsigma^\mu
-u_{-\mu}\partial_{+}\varsigma^\nu\big)
\Big].\nonumber\\
\end{eqnarray}
Varying it over $\varsigma^\mu$ and $u_{\pm\mu}$, one obtains the following equations of motion:
\begin{equation}\label{eq:emp}
\partial_{+}u_{-\mu}-\partial_{-}u_{+\mu}=0,
\end{equation}
and
\begin{eqnarray}\label{eq:jedkr}
&&\frac{\kappa}{2}\Big(\Theta^{eff}_\pm\Big)^{\mu\nu}
\big(\Delta V(\Delta U),2b\Delta \tilde{V}(\Delta U)\big)
u_{\pm\nu}
+\frac{1}{2}\partial_\pm\varsigma^\mu
=\mp\kappa(\Theta^{eff}_{0\pm}){\beta}^{\mp\mu}(V(U)),
\end{eqnarray}
where the right hand side is the contribution from the variation over the background field's arguments
\begin{eqnarray}
\delta_{A}{^\star{\cal S}}_{fix}&=&
\frac{\kappa^{2}}{2}\int d^{2}\xi
\Big[\varepsilon^{\alpha\beta}\partial_\rho (\theta^{eff})^{\mu\nu}(2b\delta\tilde{V}(U))^\rho
+
\frac{1}{2\kappa}\eta^{\alpha\beta}\partial_\rho ((G^{eff}_{E})^{-1})^{\mu\nu}\delta V^\rho(U)\Big]
\partial_\alpha U_\mu\partial_\beta U_\nu
\nonumber\\
&=&
-\kappa\int d^{2}\xi\Big[
\delta u_{+\mu}\kappa(\Theta^{eff}_{0-})^{\mu\nu}{\beta}^{+}_\nu(V(U))
+\delta u_{-\mu}\kappa(\Theta^{eff}_{0+})^{\mu\nu}{\beta}^{-}_\nu(V(U))\Big],
\end{eqnarray}
with the beta functions ${\beta}^\pm_\mu$ given by (\ref{eq:bete}).
Multiplying the equation (\ref{eq:jedkr}) by the inverse of $\Big(\Theta^{eff}_\pm\Big)^{\mu\nu}$,
$\Pi^{eff}_{\mp\mu\nu}$ defined in (\ref{eq:effpi}), we obtain the gauge fields
\begin{eqnarray}\label{eq:poljeu}
&&u_{\pm\mu}=-2\Pi^{eff}_{\mp\mu\nu}\big(\Delta V(\Delta U),2b\Delta \tilde{V}(\Delta U)\big)
\partial_\pm\varsigma^\nu
\mp2\,\beta^{\mp}_{\mu}(V(U)).
\end{eqnarray}

The gauge fixed action (\ref{eq:tdualactpom}) reduces to its initial theory (\ref{eq:tdualact}) for the equation of motion for the Lagrange multiplier
(\ref{eq:emp})
and to the effective theory for the equation of motion for the gauge fields (\ref{eq:poljeu}).
One can verify that the solution of the equation (\ref{eq:emp})
\begin{equation}\label{eq:tz}
u_{\pm\mu}=\partial_\pm\varrho_\mu,
\end{equation}
which implies $V^\mu(U)=V^\mu(\varrho)$, and $\tilde{V}^\mu(U)=\tilde{V}^\mu(\varrho)$ transforms the gauge fixed action (\ref{eq:tdualactpom}) to the T-dual action (\ref{eq:tdualact}).
 On the other hand,
substituting the gauge fields (\ref{eq:poljeu}) to the
gauge fixed action (\ref{eq:tdualactpom}),
using the zeroth order value of the gauge fields 
\begin{equation}
u_{0\pm\mu}=-2\Pi^{eff}_{0\mp\mu\nu}
\partial_\pm\varsigma^\nu=
\pm g^{E}_{\mu\nu}\partial_\pm\varsigma^\nu,
\end{equation}
which implies
\begin{equation}
U^{(0)}_\mu=g^{E}_{\mu\nu}\tilde\varsigma^\nu,
\end{equation}
while $V^{(0)\mu}(U^{(0)})=\varsigma^\mu$ and $\tilde{V}^{(0)\mu}(U^{(0)})=\tilde\varsigma^\mu$,
one obtains the effective theory (\ref{eq:aceff}), with $q^\mu=\varsigma^\mu$.
The results show 
that T-dual of the T-dual is the initial theory.

Comparing the expressions for the gauge fields (\ref{eq:tz}) and (\ref{eq:poljeu}),
we obtain the T-dual coordinate transformation law
\begin{eqnarray}\label{eq:tzakon}
\partial_\pm\varrho_\mu
&\cong&
-2\Pi^{eff}_{\mp\mu\nu}\big(\Delta V(\Delta U),2b\Delta \tilde{V}(\Delta U)\big)
\partial_\pm\varsigma^\nu
\mp2\,\beta^{\mp}_{\nu}(V(U)),
\end{eqnarray}
which is the inverse of the law (\ref{eq:zakon}).
In the zeroth order this law gives
\begin{equation}
\varrho_\mu^{(0)}\cong U^{(0)}_\mu(q).
\end{equation}
\section{Open string T-dual}
\cleq
We started with the open string described by coordinates $x^\mu$,
solved the boundary conditions and obtained the effective string described by the
even part of the initial coordinates $q^\mu$, then we T-dualized the effective theory
and obtained the T-dual string described by coordinates $\varrho_\mu$ (which were originally the Lagrange multipliers). Now,
our goal is to find an open string theory such that its effective theory,
obtained for the solution of the boundary conditions is exactly the T-dual theory (\ref{eq:tdualact}).
So, obviously the coordinates of the open T-dual string should have a lower index $y_\mu$, in order for their even part to be the T-dual coordinate $\varrho_\mu$, ones its boundary conditions are solved.
Consequently the open T-dual background should have upper indices $\widetilde{G}^{\mu\nu}$, $\widetilde{B}^{\mu\nu}$. What are the relations between the open string background and its T-dual,
and the relations between their coordinates
will become evident once the open string T-dual is found, i.e. once the connection between the effective theory of the theory we search for and the T-dual theory (\ref{eq:tdualact}) is made.

So, let us find the open string theory
\begin{equation}\label{eq:dualnao}
\widetilde{S}[y] = \kappa \int_{\Sigma} d^2\xi\
\partial_{+}y_{\mu}
\widetilde{\Pi}_{+}^{\mu\nu}(y)
\partial_{-}y_{\nu},
\end{equation}
where $\widetilde{\Pi}_{+}^{\mu\nu}=\widetilde{B}^{\mu\nu}
\pm\frac{1}{2}\widetilde{G}^{\mu\nu}$
and $\widetilde{B}^{\mu\nu}=\tilde{b}^{\mu\nu}+\frac{1}{3}\widetilde{B}^{\mu\nu\rho}y_\rho$,
such that its effective theory (\ref{eq:aceff})
\begin{eqnarray}\label{eq:deff}
\widetilde{S}^{eff} =\kappa \int d\tau \int_{-\pi}^\pi d\sigma\,
\partial_{+}q_{\mu}(y)\,
\widetilde\Pi_{+eff}^{\mu\nu}\big(q(y),2\tilde{b}\tilde{q}(y)\big)\,
\partial_{-}q_{\nu}(y),
\end{eqnarray}
is the T-dual theory (\ref{eq:tdualact}).
The effective background is composed of the metric
$(\widetilde{G}^{eff})^{\mu\nu}=\widetilde{G}_{E}^{\mu\nu}
=\big(\widetilde{G}-4\widetilde{B}^{2}\big)^{\mu\nu}$
and the Kalb-Ramond field
$B^{eff}_{\mu\nu}= -\frac{\kappa}{2}(\tilde{g}_{E} \Delta\tilde\theta \tilde{g}_{E})^{\mu\nu}$.
The effective variable is
$q_\mu(y)$, which is the even part of the variable $y_\mu$ and $\tilde{q}_\mu(y)$ is its double.
Let us first
make a connection between the variables of these two theories. We suppose that
\begin{eqnarray}\label{eq:veza}
q_\mu(y)&=&C_{\mu\nu}(g_{E}^{-1})^{\nu\rho}\tilde\varrho_\rho\,,
\nonumber\\
\bar{q}_\mu(y)&=&D_{\mu\nu}2(G^{-1}bg^{-1}_{E})^{\nu\rho}\varrho_\rho.
\end{eqnarray}
Then $\partial_\pm q_\mu(y)=\pm C_{\mu\nu}(g_{E}^{-1})^{\nu\rho}\partial_\pm\varrho_\rho$, so equating the actions (\ref{eq:deff}) and (\ref{eq:tdualact}) 
one obtains the condition for the background fields
\begin{eqnarray}\label{eq:uslovpolja}
&&g_{E}^{-1}C^{T}\widetilde\Pi_{+eff}\big(q(y),2\tilde{b}\tilde{q}(y)\big)Cg_{E}^{-1}=
-\frac{\kappa}{2}\Theta^{eff}_{-}\big(g_{E}^{-1}\tilde\varrho,2G^{-1}bg_{E}^{-1}\varrho\big).
\end{eqnarray}
In the zeroth order this condition becomes
\begin{equation}
g_{E}^{-1}C^{T}\tilde{g}_{E}Cg_{E}^{-1}=-g^{-1}_{E},
\end{equation}
which implies
\begin{eqnarray}\label{eq:gbn}
\tilde{G}&=&-(C^{T})^{-1}GC^{-1},
\nonumber\\
\tilde{b}&=&\pm(C^{T})^{-1}bC^{-1}.
\end{eqnarray}

Let us denote the variables of the T-dual theory by
\begin{equation}
z^\mu=(g_{E}^{-1})^{\mu\nu}\tilde\varrho_\nu,
\quad
 t^\mu=2(G^{-1}bg_{E}^{-1})^{\mu\nu}\varrho_\nu.
\end{equation}
Using (\ref{eq:gbn}) and (\ref{eq:veza}) we can note that
\begin{eqnarray}
\bar{q}_\mu(y)&=&2(\tilde{G}^{-1}\tilde{b})_\mu^{\ \nu}\tilde{q}_\nu(y)
\nonumber\\
&=&\mp2(CG^{-1}bC^{-1})_\mu^{\ \nu}\tilde{q}_\nu(y)
=\mp C_{\mu\nu}t^\nu,
\end{eqnarray}
on the other hand from (\ref{eq:veza}) it follows that
$
\bar{q}_\mu(y)=D_{\mu\nu}t^\nu
$, so we conclude that $D=\mp C$.
Therefore, the coordinate that solves the boundary conditions of the unknown theory (\ref{eq:dualnao}) is just
\begin{eqnarray}
y_\mu=q_\mu(y)+\bar{q}_\mu(y)=C_{\mu\nu}(z^\nu\mp t^\nu)
=C_{\mu\nu}\big((g^{-1}_{E})^{\nu\rho}\tilde{\varrho}_\rho\pm\kappa\theta_{0}^{\nu\rho}\varrho_\rho).
\end{eqnarray}

Now, we can write the first order part of the condition (\ref{eq:uslovpolja}) as
\begin{eqnarray}
g_{E}^{-1}C^{T}\tilde{G}_{E1}(Cz)Cg_{E}^{-1}&=&-(G_{E}^{-1})_{1}(z),
\nonumber\\
g_{E}^{-1}C^{T}
\tilde{g}_{E} \Delta\tilde\theta (\mp Ct)\tilde{g}_{E}
Cg_{E}^{-1}&=&\Delta\theta(t).
\end{eqnarray}
Using the explicit values of the effective open string metric, its inverse and the non-commutativity parameter
(\ref{eq:emet}),
(\ref{eq:emi}) and (\ref{eq:ncp}), we obtain
\begin{eqnarray}
\mp\big[
bC^{T}\tilde{h}(Cz)C
+C^{T}\tilde{h}(Cz)Cb
\big]
&=&
bh(z)+h(z)b,
\nonumber\\
\mp\big[
C^{T}\tilde{h}(Ct)C
+4bC^{T}\tilde{h}(Ct)Cb
\big]
&=&
h(t)+4bh(t)b,
\end{eqnarray}
with the following solution
\begin{equation}
\tilde{h}(Ct)=\mp(C^{T})^{-1}h(t)C^{-1}.
\end{equation}
Finally, we determine the open string T-dual background. It reads
\begin{eqnarray}\label{eq:dualb}
\widetilde{G}&=&-(C^{T})^{-1}GC^{-1},
\nonumber\\
\widetilde{B}(y)&=&\pm(C^{T})^{-1}(b-h(C^{-1}y))C^{-1}.
\end{eqnarray}

The effective theory (which gives the space-time equations of motion (\ref{eq:beta})) for the T-dual background (\ref{eq:dualb}) remains the same.
In a constant background, the dual background would be just (\ref{eq:gbn}),
choosing the upper solution for $\tilde{b}$ one has
\begin{equation}
\widetilde{G}\pm\tilde{b}=-(C^{T})^{-1}(G\mp b)C^{-1},
\end{equation}
which choosing $C=G\mp b$ becomes
\begin{equation}
\widetilde{G}\pm\tilde{b}=-(G\pm b)^{-1},
\end{equation}
which is in agreement with the standard T-duality relation (13) of Ref. \cite{GPV}.
The other solution would lead to equivalent conclusion.

Comparing the initial open string theory (\ref{eq:action1}) and its T-dual (\ref{eq:dualnao}) with the background (\ref{eq:dualb}),
we see that they are equal under the following transformations:
\begin{eqnarray}\label{eq:tdualnost}
\partial_\pm x^\mu&\rightarrow&\partial_\pm y_\mu,
\nonumber\\
G&\rightarrow&\widetilde{G}=-(C^{T})^{-1}GC^{-1},
\nonumber\\
B(x)=b+h(x)&\rightarrow&
\widetilde{B}(y)=\pm(C^{T})^{-1}(b-h(C^{-1}y))C^{-1}.
\end{eqnarray}

Choosing 
\begin{equation}
C_{\mu\nu}={\widetilde{G}^{-1}}_{\mu\nu},
\end{equation}
which is by (\ref{eq:dualb}) just $C_{\mu\nu}=-G_{\mu\nu}$, the T-dual of the open string in the weakly curved background becomes
\begin{eqnarray}\label{eq:tdualopen}
\widetilde{G}^{\mu\nu}&=&-(G^{-1})^{\mu\nu},
\nonumber\\
\widetilde{B}^{\mu\nu}(y)&=&\tilde{b}^{\mu\nu}+\frac{1}{3}\widetilde{B}^{\mu\nu\rho}y_\rho,
\end{eqnarray}
with the constant part of the Kalb-Ramond field equal to
\begin{equation}
\tilde{b}^{\mu\nu}=\pm\big(G^{-1}bG^{-1}\big)^{\mu\nu}
\end{equation}
and the field strength of the T-dual Kalb-Ramond field equal to
\begin{equation}
\widetilde{B}^{\mu\nu\rho}=\pm(G^{-1})^{\mu\sigma}(G^{-1})^{\nu\tau}(G^{-1})^{\rho\varepsilon}B_{\sigma\tau\varepsilon},
\end{equation}
and therefore $\widetilde{B}^{\mu\nu}(y)=\pm\big(G^{-1}B(G^{-1}y)G^{-1}\big)^{\mu\nu}$.
In this case, the transformation of the  background fields  (\ref{eq:tdualnost}) is just
\begin{eqnarray}
\Pi_{\pm\mu\nu}(x)&\rightarrow&\widetilde\Pi_\pm^{\mu\nu}(y)
=\left\{
                \begin{array}{ll}
                 \big(G^{-1}\big)^{\mu\rho}\Pi_{\mp\rho\sigma}(G^{-1}y)\big(G^{-1}\big)^{\sigma\nu}\\
                 -\big(G^{-1}\big)^{\mu\rho}\Pi_{\pm\rho\sigma}(G^{-1}y)\big(G^{-1}\big)^{\sigma\nu}\\
                \end{array}
              \right..\nonumber\\
\end{eqnarray}


\section{Conclusion}

In this paper we were looking for a T-dual of an open string moving in a weakly curved background.
The starting theory $S$, was a subject of investigation in our previous papers \cite{DS,DS1,DS2},
where it was shown that, solving the boundary conditions at the open string endpoints,
one obtains the effective closed string described by the effective closed string theory $S^{eff}$,
defined on the doubled space $(q^\mu,\tilde{q}^\mu)$.
The T-duals of such a theory can be determined using the generalized Buscher T-dualization procedure, which 
we developed earlier in Refs. \cite{DS3,DS4,DNS}. In this paper we applied the T-dualization procedure to the effective theory $S^{eff}$
along all effective directions $q^\mu$. We obtained the T-dual theory $^\star S^{eff}$. Applying the procedure
to the obtained theory along all dual directions $\varrho_\mu$, we returned to the effective theory.
So, we proved $S^{eff}\xleftrightarrow{T}{^\star S^{eff}}$.
Finally, we searched for the open string theory $\tilde{S}$ such that its effective theory is $^\star S^{eff}$ exactly. We found the explicit form of $\tilde{S}$.

The relations between the theories investigated,  
are represented in the following diagram:
\begin{center}
\noindent
\resizebox{.9\textwidth}{!}{
 \xymatrix{S = \kappa \int_{\Sigma} d^2\xi\,
\partial_{+}x^{\mu}
\Pi_{+\mu\nu}(x)
\partial_{-}x^{\nu}
 \ar[d]^{T}&\xrightarrow{solving\,BC}&
S^{eff} =\kappa \int_{\Sigma^\star} d^2\xi\, 
\partial_{+}q^{\mu}\,
\Pi^{eff}_{+\mu\nu}(q,2b\tilde{q})\,
\partial_{-}q^{\nu}
\ar[d]^{T}  
\\ 
\tilde{S}= \kappa \int_{\Sigma} d^2\xi\,
\partial_{+}y_{\mu}
\widetilde\Pi_{+}^{\mu\nu}(y)
\partial_{-}y_{\nu}
&\xrightarrow{solving\,BC}&
^\star\!{\cal S}^{eff}=\frac{\kappa^{2}}{2}\int_{\Sigma^\star} d^2\xi\, 
\partial_{+}\varrho_\mu
(\Theta^{eff}_{-})^{\mu\nu}
\big(g_{E}^{-1}\tilde\varrho,2bg_{E}^{-1}\varrho\big)
\partial_{-}\varrho_\nu.
}
}\\
\end{center}

On the left hand side are the open string theories,
the original theory depending on the original coordinate $x^\mu$
and its T-dual depending on the dual coordinate $y_\mu$. On the right hand side are the effective theories of the open string theories, obtained
for the solution of the boundary conditions. These theories are defined on the doubled spaces,
which consist of the effective variables which are the even parts of the
coordinates of the theories they originate from
and their doubles. The effective theories are T-dual to each other, and their variables are connected by the 
T-dual coordinate transformation laws (\ref{eq:zakon}) and (\ref{eq:tzakon}).
The obtained T-dual coordinate transformation laws,
could be used for investigation of the relations between the geometrical properties of the corresponding spaces. Let us notice that the T-dualization of a closed string theory \cite{DS3,DS4,DNS} led to a T-dual theory with the target space which significantly differs from the initial space. While the initial theory was defined in the ordinary space the T-dual was defined in the doubled space. In the open string case investigated here, the T-dualization does not cause such a transition. Both the initial theory and its T-dual are defined on the geometrical space.

The metrics of the above theories are
\begin{center}
\noindent
\resizebox{.6\textwidth}{!}{
\xymatrix{
G_{\mu\nu}=const\,
\ar[d]^{T}&\xrightarrow{solving\,BC}&
G^{eff}_{\mu\nu}= G^{E}_{\mu\nu}(q)\,
\ar[d]^{T}\\
\widetilde{G}^{\mu\nu}=\widetilde{const}
&\xrightarrow{solving\,BC}&
\widetilde{G}_{eff}^{\mu\nu}=(G_{E}^{-1})^{\mu\nu}(g_{E}^{-1}\tilde\varrho),
}}\\
\end{center}
and the Kalb-Ramond fields are
\begin{center}
\noindent
\resizebox{.7\textwidth}{!}{
\xymatrix{
B_{\mu\nu}(x)=b_{\mu\nu}+\frac{1}{3}B_{\mu\nu\rho}x^\rho\,
\ar[d]^{T}&\xrightarrow{solving\,BC}&
B^{eff}_{\mu\nu}= -\frac{\kappa}{2}\big(g_{E} \Delta\theta (2b{\tilde{q}})g_{E}\big)_{\!\mu\nu}\,
\ar[d]^{T}\\
\widetilde{B}^{\mu\nu}(y)=\tilde{b}^{\mu\nu}+\frac{1}{3}\widetilde{B}^{\mu\nu\rho}y_\rho
&\xrightarrow{solving\,BC}&
 \widetilde{B}_{eff}^{\mu\nu}=\frac{\kappa}{2}\theta_{eff}^{\mu\nu}(2bg_{E}^{-1}\varrho),
}}
\end{center}
with $\widetilde{G}^{\mu\nu}$ and $\widetilde{B}^{\mu\nu}(y)$  given explicitly by (\ref{eq:dualb}).

One can notice that 
the relation between the initial background and its T-dual in the open string case differs from that in the closed string case, as it should be expected.
In the closed string case, the T-duality transforms the constant metric of the weakly curved background to a coordinate dependent effective metric inverse, while the linearly coordinate dependent Kalb-Ramond field is transformed into a coordinate dependent non-commutativity parameter. In the open string case, the important role in the relation between the T-dual backgrounds plays a matrix $C$, which is introduced to define the connection between the variables of the open string theory T-dual and the effective theory T-dual. It is found that T-duality transforms the constant metric of the weakly curved background to a constant T-dual metric,
while the coordinate dependent Kalb-Ramond field transforms again to the coordinate dependent field, which is in this case of the same structure as the initial field.

\section*{Acknowledgment}

Research is supported in part by The Serbian Ministry of Education, Science and Technological Development (project No. 171031) and by The National Scholarship L'Oreal-UNESCO "For Women in Science".

\appendix

\section{Background fields}\label{sec:dod}
\cleq
The background fields used in the paper will be separated into their constant and coordinate dependent values.
\begin{itemize}
\item The weakly curved background
\begin{eqnarray}
G_{0\mu\nu}&=&const,
\nonumber\\
G_{1\mu\nu}&=&0,
\nonumber\\
B_{0\mu\nu}&=&b_{\mu\nu}=const,
\nonumber\\
B_{1\mu\nu}(x)&=&h_{\mu\nu}(x)=\frac{1}{3}B_{\mu\nu\rho}x^\rho, \quad B_{\mu\nu\rho}=const.
\nonumber\\
\end{eqnarray}
\item Effective metric $G^{eff}_{\mu\nu}(G,B)=(G_{E})_{\mu\nu}(G,B)=G_{\mu\nu}-4(BG^{-1}B)_{\mu\nu}$
\begin{eqnarray}\label{eq:emet}
G^{eff}_{0\mu\nu}=(G_{E0})_{\mu\nu}&=&g_{\mu\nu}-4b^{2}_{\mu\nu}=(g_{E})_{\mu\nu},
\nonumber\\
G^{eff}_{1\mu\nu}=(G_{E1})_{\mu\nu}&=&-4(bh+hb)_{\mu\nu}.
\end{eqnarray}
\item Effective Kalb-Ramond field $B^{eff}_{\mu\nu}$
\begin{eqnarray}
B^{eff}_{0\mu\nu}&=&0,
\nonumber\\
B^{eff}_{1\mu\nu}&=& -\frac{\kappa}{2}\Big(g_{E} \Delta\theta g_{E}\Big)_{\mu\nu}
=\big(h+4bhb\big)_{\mu\nu}.
\end{eqnarray}
Note that, because the effective Kalb-Ramond field is infinitesimal,
\begin{equation}\label{eq:effmet}
G_{E}(G^{eff}(x),B^{eff}(y))=G^{eff}(x)=G_{E}(x).
\end{equation}

\item Background field combination $\Pi^{eff}_{\pm\mu\nu}=
B^{eff}_{\mu\nu}(y)
\pm\frac{1}{2}G^{eff}_{\mu\nu}(x)$
\begin{eqnarray}\label{eq:pieff}
\Pi^{eff}_{0\pm\mu\nu}&=&\pm\frac{1}{2}(g_{E})_{\mu\nu},
\nonumber\\
\Pi^{eff}_{1\pm\mu\nu}&=&\big(h(y)+4bh(y)b\big)_{\mu\nu}
\mp2\big(bh(x)+h(x)b\big)_{\mu\nu}.\nonumber\\
\end{eqnarray}
\item Effective metric inverse $$(G_{E}^{-1})^{\mu\nu}=
\Big(G_{E0}^{-1}- 
G_{E0}^{-1}G_{E1}G_{E0}^{-1}
\Big)^{\mu\nu},$$
\begin{eqnarray}\label{eq:emi}
(G_{E}^{-1})_{0}^{\mu\nu}&=&(g_{E}^{-1})^{\mu\nu},
\nonumber\\
(G_{E}^{-1})_{1}^{\mu\nu}&=&4\Big(g_{E}^{-1}\big{(}
bh+hb\big{)}g_{E}^{-1}\Big)^{\mu\nu}.
\end{eqnarray}
\item Non-commutativity parameter $\theta^{\mu\nu}=-\frac{2}{\kappa}(G^{-1}_{E}BG^{-1})^{\mu\nu}$
\begin{eqnarray}
\theta_{0}^{\mu\nu}&=&-\frac{2}{\kappa}(g^{-1}_{E}bG^{-1})^{\mu\nu},
\nonumber\\
\theta_{1}^{\mu\nu}&=&\Delta\theta^{\mu\nu}=-\frac{2}{\kappa}\Big{[}g_{E}^{-1}(h+4bhb)g_{E}^{-1}\Big{]}^{\mu\nu}.
\end{eqnarray}

\item Effective non-commutativity parameter
\begin{eqnarray}\label{eq:ncp}
&&\theta_{eff}^{\mu\nu}:=
\theta^{\mu\nu}\big(G_{eff}(x),B_{eff}(y)\big)=
\nonumber\\
&&\qquad\,
-\frac{2}{\kappa}\Big(G^{-1}_{E}(G_{eff}(x),B_{eff}(y))B_{eff}(y)G^{-1}_{eff}(x)\Big)^{\mu\nu}
\nonumber\\
&&\theta_{0eff}^{\mu\nu}
=0,
\nonumber\\
&&\theta_{1eff}^{\mu\nu}(x,y)=
\Delta\theta^{\mu\nu}(y)
=
-\frac{2}{\kappa}\Big(
g_{E}^{-1}\big(h(y)+4bh(y)b\big)g_{E}^{-1}
\Big)^{\mu\nu}.
\end{eqnarray}
\item Tensor $\Theta^{\mu\nu}_{\pm}=-\frac{2}{\kappa}(G^{-1}_{E}\Pi_{\pm}G^{-1})^{\mu\nu}
=\theta^{\mu\nu}\mp\frac{1}{\kappa}(G^{-1}_{E})^{\mu\nu}$.
\item Effective tensor 
\begin{eqnarray}
(\Theta^{eff}_{\pm})^{\mu\nu}(x,y)&\equiv&\Theta^{\mu\nu}_{\pm}\big(G_{eff}(x),B_{eff}(y)\big)
\nonumber\\
&=&\theta^{\mu\nu}_{eff}(y)\mp\frac{1}{\kappa}(G^{-1}_{E})^{\mu\nu}(x),\nonumber
\end{eqnarray}
\begin{eqnarray}\label{eq:tetaeff}
(\Theta^{eff}_{0\pm})^{\mu\nu}(x,y)&=&
\mp\frac{1}{\kappa}(g^{-1}_{E})^{\mu\nu},
\nonumber\\
(\Theta^{eff}_{1\pm})^{\mu\nu}(x,y)&=&
\theta^{\mu\nu}_{1eff}(y)\mp\frac{1}{\kappa}(G^{-1}_{E})_{1}^{\mu\nu}(x).
\end{eqnarray}
\end{itemize}


\end{document}